\documentstyle[11pt,IAU207_pasp,twoside]{article}
\def\edcomment#1{\iffalse\marginpar{\raggedright\sl#1\/}\else\relax\fi}
\reversemarginpar

\begin{document}
\title{HST Observations of Young Star Clusters in Interacting Galaxies}
 \author{Kirk D. Borne}
\affil{Raytheon ITSS and NASA-GSFC, Code 631, Astrophysics Data Facility,
Greenbelt, MD 20771 USA}
\author{William C. Keel}
\affil{University of Alabama, Dept. of Physics and Astronomy,
Tuscaloosa, AL 35487-0324 USA}
\author{Philip N. Appleton \& Curtis Struck}
\affil{Iowa State University, Dept. of Physics and Astronomy,
Ames, IA 50011 USA}
\author{Ray A. Lucas \& Alfred B. Schultz}
\affil{STScI, 3700 San Martin Drive, Baltimore, MD 21218 USA}

\begin{abstract}
We present early results from the analysis of HST imaging
observations for several pairs of interacting galaxies.  
We include two cases that were
specifically chosen to represent a strong early (young) encounter and a weak
late (old) encounter.  The goals of the project include a determination 
of the timing, frequency, strength, and characteristics of the 
young star clusters formed in these two limiting cases of tidal
encounters.  
\end{abstract}

\section{Results from the Analysis of HST Imaging Observations}

Multi-band HST imaging data have been obtained 
for several interacting galaxy systems.  These include the Cartwheel
Ring Galaxy (Borne et al.~1996; Struck et al.~1996;
Appleton et al., in preparation), 
Arp 81 and Arp 297
(Keel \& Borne, in preparation), 
and a large sample of ULIRGs (Ultra-Luminous Infrared
Galaxies; Borne et al.~2000, and references therein).  WFPC2
$B$-band (F450W) and $I$-band (F814W) images have been obtained for
the Cartwheel, Arp 297, and Arp 81.  WFPC2 $I$-band
and NICMOS $H$-band (F160W) images have been obtained for the ULIRGs
(Borne et al.~1997b, 1999, 2001; Bushouse et al.~2001; Colina et al.~2001).
Additional HST archival WFPC2 $V$-band (F606W) images are also available for 
some of the ULIRGs (Farrah et al.~2001).
A wide variety of scientific results from these studies 
have been published or are in preparation.
Here, we report specifically on results pertaining
to the young star cluster (YSC) population in these 
interacting galaxy systems.
One of the goals of these investigations is to determine
the trigger, timing, frequency, strength, and characteristics of 
YSC formation in collisions and mergers of gas-rich galaxies.
A similar investigation by Gallagher et al.~(2001) for 
Stephan's Quintet reveals ``several 
distinct epochs of star formation that appear to trace the complex 
history of dynamical interactions in this compact group.''  
In a similar manner, we hope to trace
the history of star formation in colliding pairs.

It appears certain that the formation of YSCs in interacting galaxies 
is collision-induced (Whitmore et al.~1993, 1999; Whitmore 2000). 
A key question is:
At what phase of the interaction do the clusters form?  We attempt to
address this through investigations of a wide variety of
galaxy collisions and mergers.

A rich population of $\sim$500 very blue
YSCs is seen in the Cartwheel Ring Galaxy (Appleton et al., in preparation).
This galaxy owes its morphology to a direct head-on collision
with a companion galaxy.  The $B-I$ color image of the Cartwheel shows that
the most intensely blue colors emanate from the southwest
quadrant of the ring where the bluest and brightest YSCs are detected
(Borne et al.~1997a).
There are $\sim$150 YSCs (i.e., blue clusters)
detected in the ``spoke'' region
and in the small inner ring (a secondary by-product of
the head-on collision), 
but many more ($\sim$350) YSCs are detected in the primary
expanding collision-induced outer ring.
We have analyzed the radial color variations in pie-shaped sectors
around the Cartwheel, averaging azimuthally at a given radius within
a sector.  The azimuthal average includes the bright clusters plus 
the galactic background and is generally redder than the clusters,
probably because there is some underlying galaxy light which is red and
centrally concentrated, with a rough $1/r$ distribution.
The azimuthally averaged radial color distribution
and the colors of the YSCs converge to very blue values of $B-I$
at large radius where the YSCs dominate the light of the galaxy
(in the outer ring).

While the trigger and timing of YSC formation is clear in 
the case of the Cartwheel galaxy, there is
still some uncertainty as to when the YSCs begin to form in grazing
encounters.  We are investigating two extremes of the population
of such collision configurations: a strong/young encounter (Arp 81) 
and a weak/old (well developed) encounter (Arp 297). 
Objects that appear to be YSCs are seen in both cases.

In Arp 297 (NGC 5752/5754), nearly 800 star clusters have been identified,
mostly in the larger galaxy (NGC 5754).  The latter are
much fainter and yet bluer (on average) compared to the
clusters in the small companion galaxy (NGC 5752).  Thus, this
weak/old tidal interaction
has stimulated the most luminous clusters to form in the small galaxy,
which feels the strongest tidal perturbation, and these clusters 
probably formed earlier in the interaction (hence their redder colors)
compared to the less luminous clusters in NGC 5754,
which probably formed much later in the interaction, 
therefore appearing younger (bluer colors).

The true-color HST image of Arp 81 (NGC 6621/6622) reveals a
remarkable pattern of dust lanes and filaments.
Many ($\sim$700) luminous blue clusters have been identified,
clearly demonstrating YSC formation in a young, yet strong, collision.
The most populous region for blue clusters in Arp 81 is
the contact region between the two galaxies, reminiscent of
a similarly intense starburst event in the contact region
of the Antennae (NGC 4038/4039; Whitmore et al.~1999).
The colors of the YSCs in Arp 81 are intermediate between those 
of the two galaxies in the Arp 297 galaxy pair (described above),
and hence the YSCs in Arp 81 may be of a corresponding intermediate age.

In addition to these results, we find that
the ULIRGs show a significant population of very bright
star knots or YSCs (Borne et al.~1997b; Surace et al.~1998), which are
most likely produced during the starburst/merger event.  In one case
(IRAS~21130$-$4446), HST's superb angular
resolution capability has revealed 
that this ULIRG is a collision-induced ring galaxy
with a ``beads on a string'' morphology 
(Fig.~2b; Borne et al.~1997b).  These ``beads''
are probably very bright collision-induced YSCs, similar to (though
much brighter than) those seen in the Cartwheel.  
This ring galaxy may prove to be very useful 
as a comparative chronometer of the collision age, the ULIRG 
development age, and the YSC formation age.

{\bf{Acknowledgments.}}
This work was supported by a Raytheon Sabbatical Research
Award and by 
NASA through HST grants 
GO-05410.01-93, GO-06346.01-95A, and GO-07467.02-96A
from the Space Telescope Science Institute,
which is operated by AURA, Inc., under NASA contract NAS5-26555.

\end{document}